\documentclass[journal]{IEEEtran}
\usepackage{amsmath}
\usepackage{amssymb}
\ifCLASSINFOpdf
\else
   \usepackage[dvips]{graphicx}
\fi
\usepackage{url}
\usepackage[ruled,vlined]{algorithm2e}
\usepackage{graphicx}

\begin{document}

\title{ Adaptive User Pairing for Downlink NOMA System with Imperfect SIC}

\author{Nemalidinne Siva Mouni, \IEEEmembership{Student Member, IEEE}, Abhinav Kumar, \IEEEmembership{Senior Member, IEEE} and Prabhat K. Upadhyay, \IEEEmembership{Senior Member, IEEE }}



\maketitle

\begin{abstract}
Non-orthogonal multiple access (NOMA) has been recognized as a key driving technology for the fifth generation (5G) and beyond 5G cellular networks. For a practical dowlink NOMA system with imperfect successive interference cancellation (SIC), we derive bounds on channel coefficients and power allocation factors between NOMA users to achieve higher rates than an equivalent orthogonal multiple access (OMA) system. We propose an adaptive user pairing (A-UP) algorithm for NOMA systems. Through extensive simulations, we show that NOMA with imperfect SIC is not always superior to OMA. Further, the proposed A-UP algorithm results in better performance than state-of-the-art NOMA pairing algorithms in presence of SIC imperfections.

\end{abstract}

\begin{IEEEkeywords}
Achievable sum rate (ASR), imperfect successive interference cancellation (SIC), non-orthogonal multiple access (NOMA), user pairing.
\end{IEEEkeywords}

\IEEEpeerreviewmaketitle

\section{Introduction}
\IEEEPARstart{O}{ne} of the key technologies for the fifth generation (5G) and beyond 5G cellular networks is non-orthogonal multiple access (NOMA) \cite{MAT}. In NOMA, symbols intended for multiple users are multiplexed in the power domain at the transmitter in the same space-time-frequency resource. The optimal user pairing and power allocation between paired users are essential to maximize the full potential of NOMA. Conventionally, users with larger channel gain differences have been paired in NOMA to achieve rates greater than OMA (orthogonal multiple access) as illustrated in \cite{cha_diff}. The uniform channel gain difference (UCGD) pairing algorithm has been proposed to overcome the mid users pairing issues in conventional near-far user NOMA pairing schemes \cite{pair_main}. In \cite{angle1,angle2}, novel angle-based NOMA pairing algorithms have been proposed by taking directionality into account.

In NOMA, successive interference cancellation (SIC) is performed at the receiver of the higher channel gain user to remove the signal of the other user in the pair. It has been shown in \cite{SIC} that imperfect SIC impacts the outage and throughput performance of the NOMA users. However, most of the existing user pairing schemes do not consider the effect of imperfect SIC in user pairing. Few works exist which have taken into account the imperfect SIC, such as the hierarchical distributed power bandwidth allocation (PBA) scheme in \cite{pair_5}. To the best of our knowledge, there is no work directed on the trade-off between imperfect SIC NOMA and OMA till date.

\begin{figure}[htp]
\includegraphics[width=0.5\textwidth ]{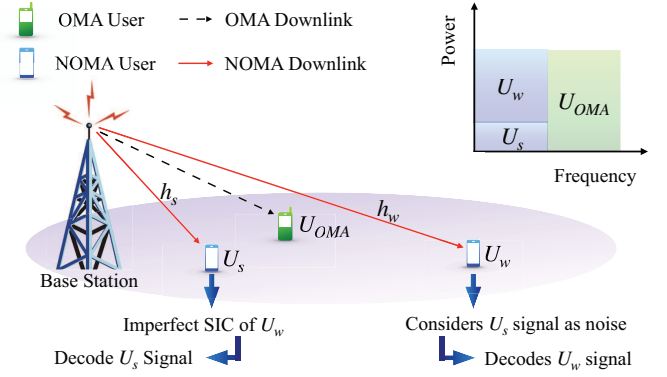}
\caption{ System Model }
\label{fig:figure3}
\end{figure}

This letter presents the first work that derives bounds on user channel gains based on which they can either be paired as NOMA or considered as OMA users in presence of imperfect SIC. We propose a novel adaptive user pairing (A-UP) scheme based on minimum signal-to-interference-plus-noise ratio (SINR) difference criteria for two users in a NOMA pair. Bounds are also derived on power allocation between the users paired through A-UP to maximize their achievable sum rate (ASR).

The organization of the paper is as follows. The system model is presented in Section II. In Section III, the various bounds are derived. The proposed A-UP algorithm is described in Section IV. Numerical results utilizing both log rate (LR) and discrete rate (DR) model are presented in Section V. Section VI comprises of some concluding remarks and possible future works.

\section{System Model}
We consider a 5G cellular network and focus on the downlink NOMA pairing for the base station (BS) under consideration as depicted in Fig. 1. Given any user association scheme, we presume the set of users connected to the BS is given by $U=\{1,2,\ldots,N\}$. The downlink SINR from the transmitter (i.e., the BS) to a receiver (user $u\in U$), on a subchannel in case of OMA system is formulated as
\begin{equation}
\gamma_{u} = P_t \frac{|h_u|^2}{N_0 + I},
\end{equation}
where $P_t$ is the transmitted power, $h_u$ is the channel gain of user $u$, $N_0$ the variance of additive white Gaussian noise, and $I$ the interference from nearby BSs.
Given a LR model in an OMA system, the normalized downlink rate for a user $u$ can be expressed as
\begin{equation}
 R_{u}^{\textmd{\tiny{OMA}}} =  \frac{1}{2}\log\left (1 + \gamma_{u} \right).
\end{equation}
where the factor ${1}/{2}$ is considered owing to the loss in multiplexing in OMA system.

For NOMA LR formulation, we pick a strong user $U_s$ and a weak user $U_w$ having a significant channel gain difference (i.e., $|h_s|^2 > |h_w|^2$) from the BS under consideration. Let $\alpha_s$ and $(1-\alpha_s)$ be the fraction of transmitted power allocated to $U_s$ and $U_w$, respectively. The SINR of the NOMA user pair i.e., $\hat{\gamma}_{s}$ and $\hat{\gamma}_{w}$ can be, respectively, expressed as
\begin{align}
    \hat{\gamma}_{s} &=  \frac{\alpha_{s}P_t |h_s|^2}{ N_0 + I + \beta(1-\alpha_s)P_t |h_s|^2 } \nonumber \, \mbox{ and} \\
    \hat{\gamma}_{w} &= \frac {(1-\alpha_{s})P_t |h_w|^2}{N_0 + I + \alpha_{s}P_t |h_w|^2},
\end{align}
where $\beta \in [0,1]$ represents the imperfection in SIC. A value of $ \beta=0 $ implies that the strong user is completely able to remove the interference from weak user, i.e., perfect SIC. Assuming $\gamma_s$ and $\gamma_w$ to be the SINR values of strong and weak users in case of OMA, then using (1), we can reformulate (3) as follows
\begin{align}
\hat{\gamma}_{s} &= \frac{\alpha_{s}\gamma_s}{ 1 + \beta(1-\alpha_s)
\gamma_s},\nonumber \\
\hat{\gamma}_{w}& =  \frac {(1-\alpha_{s})\gamma_w}{1 + \alpha_{s}\gamma_w}.
\end{align}

The SINR values obtained in (4) can be used to compute the NOMA rates, i.e., $R_{s}^{\textmd{\tiny{NOMA}}}$ and $R_{w}^{\textmd{\tiny{NOMA}}}$, respectively, as 
\begin{gather}
   R_{s}^{\textmd{\tiny{NOMA}}} =  \log \left (1 + \hat{\gamma}_{s} \right) \nonumber \mbox{ and}\\
    R_{w}^{\textmd{\tiny{NOMA}}} =  \log \left (1 + \hat{\gamma}_{w} \right).
\end{gather}

\noindent Further, the ASR of NOMA pair is obtained by using (5) as
\begin{equation*}
  ASR^{\textmd{\tiny{NOMA}}} =  R_{s}^{\textmd{\tiny{NOMA}}} + R_{w}^{\textmd{\tiny{NOMA}}}.
\end{equation*}
Similarly, the ASR of OMA pair is 
\begin{equation*}
  ASR^{\textmd{\tiny{OMA}}} =  R_{s}^{\textmd{\tiny{OMA}}} + R_{w}^{\textmd{\tiny{OMA}}}.
\end{equation*}
\section{Computation of Bounds  }
In this section, we derive bounds on channel coefficients as well as power allocation for achieving better NOMA rates as compared to OMA.

\subsection{Bounds on $\alpha_s$}
\subsubsection{Considering weak user's SINR}
For the upper bound on $\alpha_s$, we consider the constraint that the rate of weak user in NOMA should be greater than its OMA rate ($R_{w}^{\textmd{\tiny{NOMA}}} > R_{w}^{\textmd{\tiny{OMA}}}$). The same can be expressed as
\begin{equation}
\log \left (1 + \frac {(1-\alpha_s)\gamma_w}{1 + \alpha_s \gamma_w }\right ) > \frac{1}{2}\log \left(1 + \gamma_w \right ).
\end{equation}

\noindent The inequality in (6) can be simplified to obtain an upper bound on $\alpha_s$ as
\begin{equation}
\alpha_{s} < \frac{1}{\gamma_w}\big(\sqrt{1 + \gamma_w} - 1 \big) \triangleq \alpha_s^*.
\end{equation}

The inequality in (7) states that in case $\alpha_s > \alpha_{s}^*$, the weak user rate in NOMA is less than the corresponding OMA system. Further, we observe that $\alpha_{s}^*$ is monotonically decreasing function with respect to (w.r.t.) $\gamma_w$ ($\gamma_w > 0$), with a maximum value being 0.5. Thus, $0 < \alpha_s < 0.5$.

\subsubsection{Considering strong user's SINR}
We impose a similar constraint on strong user rate, i.e., $R_{s}^{\textmd{\tiny{NOMA}}} > R_{s}^{\textmd{\tiny{OMA}}}$ which results in
\begin{equation}
 \log \left (1 + \frac{\alpha_s \gamma_s}{ 1 + \beta(1-\alpha_s) \gamma_s} \right ) > \frac{1}{2}\log \left (1 + \gamma_s\right ).
\end{equation}

\noindent On solving for $\alpha_s $, the inequality in (8) reduces to
\begin{equation}
\alpha_s > \frac{(1 + \beta \gamma_s)(\sqrt{1 + \gamma_s} -  1)}{ \gamma_s(1 + \beta \sqrt{1+\gamma_s} - \beta)},
\end{equation}

\noindent where (9) gives a lower bound on power allocation for a given value of $\beta$. As long as this bound is satisfied, the NOMA based rate of strong user exceeds the OMA rate.

It is evident from (7) and (9) that there exists a trade-off between strong and weak user's rate w.r.t. $\alpha_s$. Thus, $\alpha_s$ has to be carefully selected to ensure higher individual NOMA rates for strong and weak users as compared to their OMA rates. Further, the presence of $\beta$ in (9) may result in scenarios where $ASR^{\textmd{\tiny{NOMA}}}$ is lower than $ASR^{\textmd{\tiny{OMA}}}$. Hence, we next study the impact of $\beta$ on these rates.

\subsection{Imperfect SIC parameter }

Considering the constraint of $ASR^{\textmd{\tiny{NOMA}}} > ASR^{\textmd{\tiny{OMA}}}$ with $\beta \in (0,1)$ results in

 \begin{equation}
\beta < \frac{(1 + \alpha_s \gamma_s)\sqrt{1 + \gamma_w} - (1 + \alpha_s \gamma_w)\sqrt{1 + \gamma_s}}{ \gamma_s(1-\alpha_s)(\sqrt{1 + \gamma_s}(1 + \alpha_s \gamma_w) - \sqrt{1 + \gamma_w})}.
\end{equation}

\noindent Substituting the upper bound of $\alpha_s$ from (7) in (10), we obtain

\begin{equation}
\beta < \frac{\gamma_w - \gamma_s  + \gamma_s\sqrt{1+\gamma_w} - \gamma_w\sqrt{1 + \gamma_s} }{\gamma_s(\sqrt{1 + \gamma_s} - 1)(\gamma_w  - \sqrt{1 + \gamma_w} +1)}.
\end{equation}

The expression in (11) states that for any user pair if imperfect SIC parameter is within the upper bound, then, all users in a NOMA pair will have more rate via NOMA as compared to OMA. Note that $ASR^{\textmd{\tiny{NOMA}}}$, $R_{w}^{\textmd{\tiny{NOMA}}}$ greater than $ASR^{\textmd{\tiny{OMA}}}$, $R_{w}^{\textmd{\tiny{OMA}}}$, respectively, need not imply $R_{s}^{\textmd{\tiny{NOMA}}} > R_{s}^{\textmd{\tiny{OMA}}}$.

It can be observed from (10), that for certain values of $\alpha_s$, the numerator becomes negative (denominator is always positive). Applying positivity constraint on the numerator, we get

\begin{equation}
\alpha_s > \frac{1}{\sqrt{1+ \gamma_s } + \frac{1}{\sqrt{1 + \gamma_w}}}.
\end{equation}

For any given $\gamma_s$ and $\gamma_w$, it can be observed that upper bound in (7) is always greater than lower bound in (9) provided $\gamma_s > \gamma_w$ and $\beta = 0$. However, this is not the same case with (12). The lower bound on $\alpha_s$ from (12) exceeding the upper bound in (7) implies that even if $ASR^{\textmd{\tiny{NOMA}}} > ASR^{\textmd{\tiny{OMA}}}$ both the individual user rates may not exceed corresponding individual OMA rates.

\subsection{Minimum SINR Difference (MSD) }
The MSD required between two users in a particular NOMA system to be paired is expressed as $\Delta_{MSD} $. For this, we take into consideration that the upper bound in (7) should always be greater than the lower bound in (12) resulting in :

\begin{equation}
\gamma_w < \frac{(\sqrt{1 + \gamma_w} - 1)(\sqrt{1 + \gamma_s}\sqrt{1 + \gamma_w} + 1)}{\sqrt{1 + \gamma_w}}.
\end{equation}

\noindent On subtracting $\gamma_s$ on both sides of (13) and rearranging the above equation results in :

\begin{equation}
\gamma_s  - \gamma_w > \gamma_s - \frac{(\sqrt{1 + \gamma_w} - 1)(\sqrt{1 + \gamma_s}\sqrt{1 + \gamma_w} + 1)}{\sqrt{1 + \gamma_w}}.
\end{equation}

\noindent Thus, the MSD is given as
\begin{equation}
\Delta_{MSD} = \gamma_s - \frac{(\sqrt{1 + \gamma_w} - 1)(\sqrt{1 + \gamma_s}\sqrt{1 + \gamma_w} + 1)}{\sqrt{1 + \gamma_w}}.
\end{equation}

For a given pair of two users, if $\gamma_{s} - \gamma_{w} > \Delta_{MSD}$ then we can pair them as NOMA users. However, if this constraint in (15) is not satisfied, then pairing such users may not be beneficial. Next, we utilize the MSD criterion in (15) for the adaptive NOMA user pairing algorithm.
\section{Adaptive User Pairing (A-UP)}

In existing works like \cite{cha_diff, pair_main}, typically, the sum rate of NOMA pairs has been maximized. In case of perfect SIC, this sum rate maximization results exceed OMA rates. However, as shown in the previous section, imperfect SIC can result in OMA rates being higher than NOMA rates for certain scenarios. Further, even if the ASR in NOMA is greater than OMA, it need not ensure that both users in the NOMA pair strictly achieve NOMA rates greater than OMA rates. This can occur because sum rate maximization may result in most of the power being allocated to the strong user resulting in higher sum and strong user NOMA rates at the cost of the weak user. Few papers have considered a threshold for both the user rates \cite{pair_main}. However, no mechanism with a theoretical foundation has been presented for this threshold selection. Motivated by the bounds derived in the previous section, we present the following NOMA user pairing algorithm.

We only perform the NOMA user pairing for users that satisfy the criterion in (15). Users that do not satisfy (15) with any other user are designated as OMA users. We present an MSD based A-UP algorithm which aims to maximize the sum rate. We illustrate the algorithm with the following example. Let us consider $N = 8$ users. We calculate the SINR values for given $P_t$, $N_0$ and the interference from the BS, and split them into two groups such that the first half remains in group $G_1$ and the rest in group $G_2$. We re-arrange the group $G_1$ in descending order and stack them adjacent to each other as in Fig. 2, such that now both the groups have mid users on top of the stack.

\begin{figure}[htp]
\includegraphics[width=0.5\textwidth]{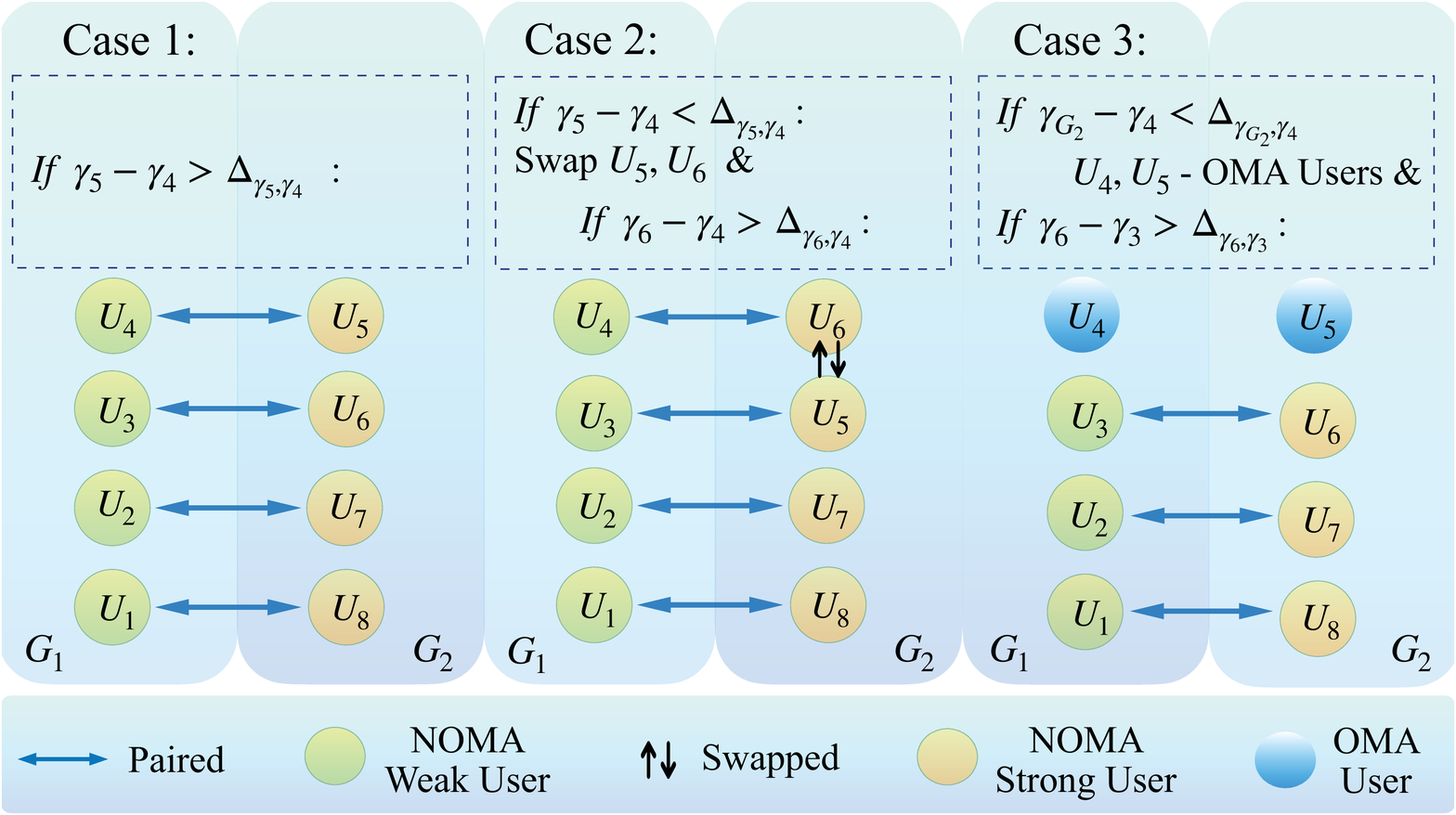}
\caption{ Illustration of the proposed pairing scheme. }
\label{fig:figure3}
\end{figure}

\begin{algorithm}
 \SetAlgoLined

  INPUTS : $ G_1, G_2, P_t, N_o$ \newline
 1. Initialize $i =1, k1 = N/2, k2 = N/2 +1 $ \newline
 2. Set $j=1 $  \newline
 3. \textbf{While} ( $k2 \leq N $)\newline
 4. \hspace{0.5 cm} \textbf{if} ($j \leq N/2$) \newline
 5. \hspace{1 cm} Select $\gamma_{k1}$ from $G_1$ and $\gamma_{k2}$ from $G_2$ \newline
 6.  \hspace{1.5 cm} \textbf{if} ($\gamma_{k2} - \gamma_{k1} > \Delta_{MSD}$ ) \newline
 7. \hspace{1.8 cm} $U_{k1},U_{k2}$ will be paired; \newline
 8. \hspace{1.8 cm} $ i = i+1; j = i ; $ \newline
 9. \hspace{1.8 cm} $k1 = k1 -1 ; k2 = k2 + 1$, \newline
 10.\hspace{1.4 cm} \textbf{elseif} ($k2 == N$) \newline
 11.\hspace{1.8 cm} $ k2 = N/2 + i ;i = i+1; j = i ; $ \newline
 12.\hspace{1.8 cm} $U_{k1}$ and $U_{k2}$ will be OMA users, \newline
 13.\hspace{1.8 cm} $k1 = k1 -1 ; k2 = k2 + 1$, \newline
 14.\hspace{1.8 cm} Move to step 4 \newline
 15.\hspace{1.4 cm} \textbf{else}\newline
 16.\hspace{1.8 cm} $j=j+1, k2 = k2 +1$; \newline
 17.\hspace{1.4 cm} \textbf{end} \newline
 18.\hspace{0.5 cm} \textbf{end}\newline
 \caption{\!Adaptive User Pairing (A-UP)\! Algorithm}
\end{algorithm}

Now, we compute the  $\Delta_{MSD}$ as per (15) for the top two users in $G_1$ and $G_2$, to check the MSD criteria. If the MSD criteria is satisfied in case of first pair, i.e., mid pair then we can directly pair the consecutive users in both the groups as shown in Case 1 of Fig. 2. This is because the SINR values in both the groups are in ascending order, ensuring maximum SINR difference.

Case 2 of Fig. 2 presents the instance of MSD not being satisfied for the first user pair from both the groups. In such a case, we swap the current user under consideration in $G_2$ with the next user. That is to bring in a user with more SINR value to top of the stack in $G_2$ which may satisfy the MSD condition with the user under consideration in $G_1$. Once the condition is met, we pair those users and check the MSD for next set of users.

\begin{figure}[htp]
\includegraphics[width=0.5\textwidth]{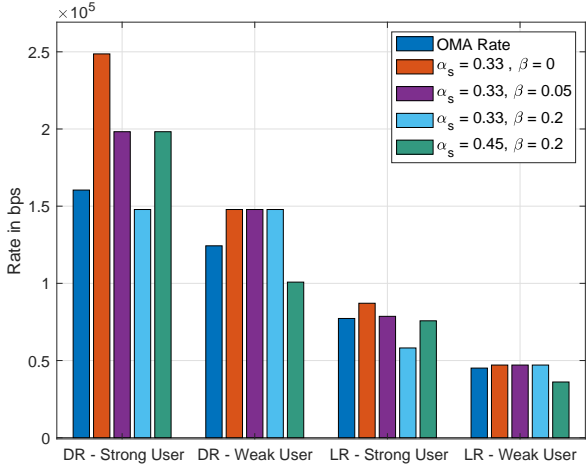}
\caption{Illustration of rate trade off between strong and weak user.    }
\label{fig:figure3}
\end{figure}

In Case 2 of Fig. 2, we swapped the users $U_5$, $U_6$ and this satisfied the MSD in case of $U_4$ and $U_6$, thus forming the pair. Further, the consecutive users from both the groups have been paired as shown in Case 2 of Fig. 2, as they satisfy the MSD condition. Following the same procedure, it may happen that this MSD criteria is not satisfied for a particular user from $G_1$ with any user in $G_2$. In such a case, it will remain to be an OMA user. The same is presented in Case 3 of Fig. 2, where $U_4$ and $U_5$ will be OMA users. It may happen that $U_3$ and $U_5$ are able to achieve the required criteria, but pairing them so leads to pairing of next consecutive users leaving $U_8$ unpaired. $U_8$ having the maximum channel gain will receive maximum rate if paired and conventionally also we try to pair the near-far users most of the time. That is why, $U_5$ is left unpaired along with $U_4$.

Considering this approach, a heuristic approach to pair the users adaptively considering the MSD has been presented in Algo. 1. Next, we present the numerical results.

\section{Numerical Results}

The simulations are carried out for poisson point distributed BSs and user positions with densities $25$ BS/km$^{2}$ and $120$ users/km$^{2}$, respectively, as in \cite{bs}. For the pathloss model and the simulation parameters, we consider an urban cellular environment from \cite{std} along with Rayleigh fading model (scaling parameter $ = 1$) and omni-directional antennas. Based on the maximum SINR value of each user received from all the BSs, we assign the user to a particular BS. Further, we pick two users from a BS which have significant channel gain difference to study the variation of $\alpha_s$ and $\beta$.

Fig. 3 compares the NOMA rates of strong and weak users with their OMA rates considering both perfect as well as imperfect SIC. For $\alpha_s = 0.33$ satisfying the upper bound as well as lower bound for the pair under consideration, i.e., $\gamma_s = 8.64 $ dB and $\gamma_w = 3.88 $ dB, we see that there is significant increase in strong as well as weak user's rate when compared to the corresponding OMA rates. For the same $\alpha_s = 0.33 $ but imperfect SIC parameter $\beta = 0.05$, we observe that the strong user NOMA rate drops now in both LR and DR models. Still the NOMA rates are greater than the OMA owing to the fact that $\beta = 0.05$ is within the upper bound given by (11) for $\gamma_s = 8.64 $ dB and $\gamma_w = 3.88 $ dB. If we increase $\beta$ to 0.2, i.e., much greater than the upper bound, strong user rate falls below its OMA rate.  Nevertheless, weak user rate remains same as it doesn't depend on $\beta$. However, if we try to increase the NOMA rate of strong user by increasing $\alpha_s$ to 0.45, weak user rate drops. Thus, we can say that there is always a trade off between the strong and weak user rate for different values of $\alpha_s$, when imperfect SIC is considered.

\begin{figure}[htp]
\includegraphics[width=0.5\textwidth]{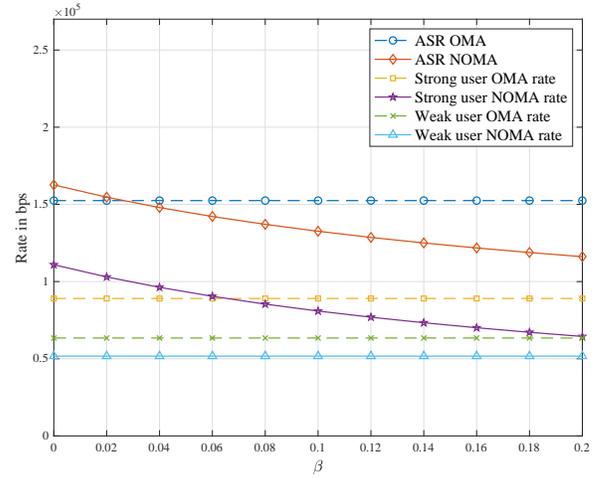}
\caption{ Variation of $\beta$ for a fixed value of $\alpha_s = 0.32$: LR model.   }
\label{fig:figure3}
\end{figure}

\begin{figure}[htp]
\includegraphics[width=0.5\textwidth]{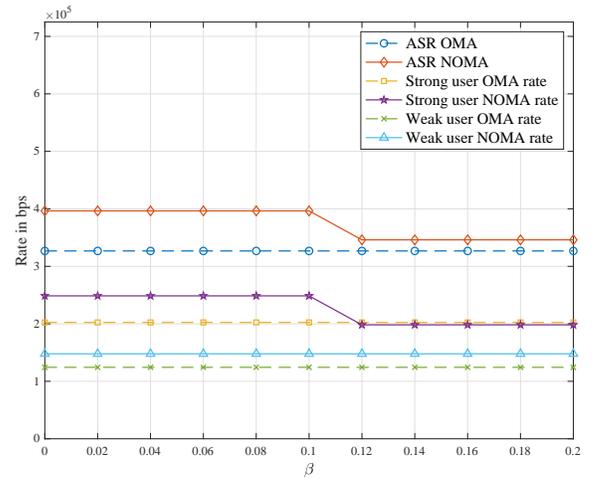}
\caption{ Variation of $\beta$ for a fixed value of $\alpha_s = 0.32$: DR model.   }
\label{fig:figure3}
\end{figure}

The effect of variation of $\beta$ while computing the rates in LR model and DR model is shown in Fig. 4 and Fig. 5, respectively, for $\gamma_s = 10.48 $ dB, $\gamma_w = 4.69 $ dB and $\alpha_s = 0.32$. This value of $\alpha_s$, i.e., 0.32 is within the upper and lower bounds as per (7) and (12). For the same users, the upper bound computed for $\beta$ as per (11) is 0.06. This is in accord with the simulated results in both LR and DR models.

\begin{figure}[htp]
\includegraphics[width=0.5\textwidth]{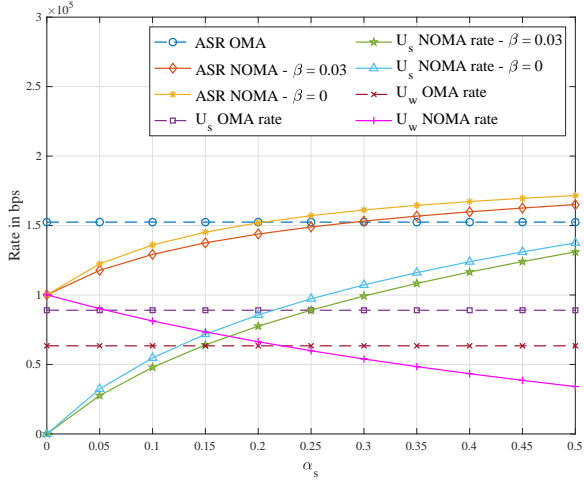}
\caption{ Variation of $\alpha_s$ for a fixed value of $\beta = 0.02$: LR model.   }
\label{fig:figure3}
\end{figure}

\begin{figure}[htp]
\includegraphics[width=0.5\textwidth]{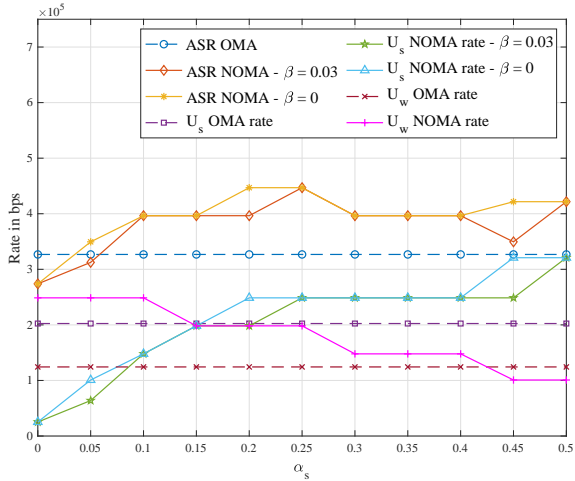}
\caption{ Variation of $\alpha_s$ for a fixed value of $\beta = 0.02$: DR model.   }
\label{fig:figure3}
\end{figure}

Fig. 6 and Fig. 7 illustrate the variation of $\alpha_s$ in case of perfect SIC ($\beta = 0$) and imperfect SIC ($\beta = 0.02$), considering the same users, i.e., $\gamma_s = 10.48 $ dB and $\gamma_w =  4.69 $ dB in LR and DR models. The impact of $\beta$ on strong user is visible in both the plots, wherein, the rate of strong user is more in case of perfect SIC when compared with the imperfect SIC. We observe that at the lower bound of $\alpha_s$, i.e., 0.25 as per (12), NOMA rate of strong user exceeds OMA. Moreover, beyond the upper bound, i.e., $\alpha_s = 0.33$ as per (7), OMA rate of weak user exceeds NOMA rate.

The performance of the proposed algorithm has been compared with the conventional near-far (NF) and UCGD pairing algorithms in Fig. 8, presuming $\beta = 0.13$. The results have been averaged over 80 independent Rayleigh fading realizations. OMA based ASR has also been considered, which is greater than NOMA ASR due to the imperfect SIC by NOMA users as in Fig. 8. It is apparent from Fig. 8 that the proposed algorithm is efficient in achieving higher ASR as compared to the state-of-the-art NOMA user pairing algorithms. This is because, the A-UP algorithm works as trade off between the purely OMA and purely NOMA users.

\begin{figure}[htp]
\includegraphics[width=0.5\textwidth]{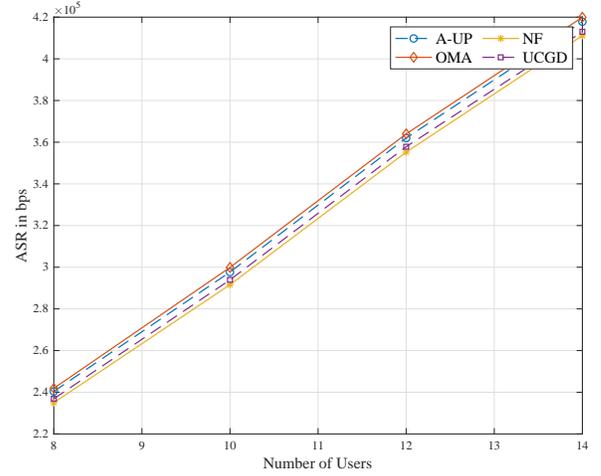}
\caption{Comparative ASR performance between A-UP, OMA, NF and UCGD for different number of users given $\beta = 0.13$. }
\label{fig:figure3}
\end{figure}

\section{Conclusion }
We have shown that while trying to maximize the total ASR in a NOMA system, there is always a trade off between the user rates achieved by the strong and weak users in case of imperfect SIC. We have derived conditions in which not utilizing NOMA for certain users may be beneficial for the overall system. Further, optimal bounds on fraction of power to be divided between NOMA user pairs and the imperfect SIC have been derived to ensure NOMA rates are better than OMA rates. An adaptive NOMA user pairing algorithm has been presented that performs better than state-of-the-art works in presence of imperfect SIC. In future, we plan to test the derived bounds and proposed algorithm on a hardware testbed.

\section{Acknowledgements}
This work was supported in part by the project BRICS: Design and Development of Large-Scale Ambient Energy Harvesting Wireless Networks (LargEWiN).

\end{document}